\documentclass{aa}
\usepackage{epsfig}
\usepackage{rotating}
\usepackage{txfonts}
\usepackage{aalongtable,lscape}
\newcommand{\gesssim}{\mathrel{\hbox{\rlap{\hbox{\lower4pt\hbox{$\sim$}}}\hbox{$>$}}}}

\newcommand{\teff}{$T_{\rm eff}$}

\begin{document}

\title{Element abundances
in the metal rich open cluster NGC~6253 \thanks{Based on observations
collected at ESO telescopes under program 072.D-0550(A)} \thanks{Table 3 
is only available in electronic form
at the CDS via anonymous ftp to cdsarc.u-strasbg.fr (130.79.128.5)
or via http://cdsweb.u-strasbg.fr/cgi-bin/qcat?J/A+A/}}

   \subtitle{}

   \author{P. Sestito\inst{1,2} \and S.
Randich\inst{1} \and A. Bragaglia\inst{3}}

   \offprints{P. Sestito}

\institute{INAF-Osservatorio Astrofisico di Arcetri, Largo E.~Fermi 5,
             I-50125 Firenze, Italy\\
\email{sestito, randich@arcetri.astro.it}
\and
INAF-Osservatorio Astronomico ``G.S. Vaiana'' di Palermo, Piazza del Parlamento 1, I-90134 Palermo, Italy
\and
INAF-Osservatorio Astronomico di Bologna, Via C. Ranzani 1,
            I-40127 Bologna, Italy\\
\email{angela.bragaglia@oabo.inaf.it}
            }

\titlerunning{Abundances in NGC~6253}
\date{Received Date: Accepted Date}

  \abstract
   {We have carried out a big FLAMES survey of 10 Galactic open clusters aiming at
different goals. One of them is the determination of chemical abundances, in order
to put constraints on the radial metallicity gradient in the disk
and its evolution.
One of the sample clusters is the very metal rich NGC 6253.}
{We have obtained UVES high resolution spectra
of seven candidate cluster 
members (from the turn off up to the red
  clump) with the goal of determining the chemical composition
of NGC 6253 and to investigate
its origin and role in the interpretation of
the radial metallicity gradient in the disk.}
{Equivalent width analysis and spectral synthesis
were performed using MOOG and Kurucz model atmospheres.}
{We derived abundances of Fe, $\alpha$- and Fe-peak elements, the
  light element Na and the s-process element Ba.
Excluding two likely non-members and the clump giant, whose
metallicity from equivalent widths is overestimated,
we find an average [Fe/H]=+0.36$\pm$0.07 (rms)
for the cluster. For most of the 
other elements we derive
solar abundance ratios.}
{}

\keywords{ Stars: abundances --
           Stars: Evolution --
           Galaxy:disk --
           Open Clusters and Associations: Individual: NGC~6253 }
\maketitle

\section{Introduction}\label{intro}

The advent of new observational capabilities in the last years allowed
a steady improvement 
in the field of chemical abundances in astrophysical objects 
studied through high resolution spectroscopy.
In particular, it is now possible to derive precise element abundances
in old and distant Galactic open clusters, allowing us
to address different issues such as
the origin of the clusters themselves, the Galactic
radial metallicity gradient,
and the formation and evolution of the Galactic disk 
(see, e.g. Friel \cite{friel06} and references therein).
At the
same time, the abundances of other
species, such as $\alpha$- and Fe-peak elements, or
the s- and r-process elements, and their ratios to
Fe, are crucial to get insights on the role of stars with different masses and
evolutionary lifetimes in the heavy element enrichment of the interstellar
medium. This,
through comparison with Galactic
enrichment models, permits to put constraints
on the initial mass function and star formation history
during the early phases of disk evolution.

In this context, we carried out a VLT/FLAMES program
on a sample of 10 open clusters (Randich et al.~\cite{messenger}).
One of the main goals of this project is the determination
of the cluster metallicity and chemical composition
through the analysis of UVES spectra of evolved members (Sestito
et al. \cite{sestito06} -- hereafter
Paper~{\sc i}).
We focus here on the $\sim$3 Gyr old cluster
\object{NGC 6253}; this object, located
towards the Galactic center, is one of the most interesting
clusters
in our sample, since it has a metallicity considerably higher than solar.
Twarog, Anthony-Twarog, \& de Lee (\cite{ttl03})
suggested that
this cluster might be the most metal-rich object in the Galaxy.
From photometric indices,
they found
[Fe/H]$\sim$+0.7, while from comparison with isochrones they concluded
that
$\alpha$-enhanced 
isochrones provided the best fit, indicating a  [Fe/H] closer to +0.4.
Besides the study by Twarog et al.,
various photometric surveys of this cluster were carried out
in the last $\sim$10 years
(Bragaglia et al.~\cite{bragaglia97};
Piatti et al.~\cite{piatti}; Sagar et al.~\cite{sagar});
most of them yielded 
very similar values of reddening, age, and distance,
namely
$E(B-V)\simeq0.20$, age $\sim$3 Gyr,
and $(m-M)_{0}=10.9-11.0$
($d\sim1.8$ kpc).

The first
spectroscopic determination of the metallicity of NGC~6253 
was carried out by Carretta
et al.~(\cite{carretta00}), who found [Fe/H]=+0.36$\pm$0.20.
A more recent analysis
(Carretta, Bragaglia,
\& Gratton 2007, submitted) based on better quality spectra
of five red clump stars favours a higher metallicity
([Fe/H]=+0.46).

Besides NGC~6253, a few other open clusters with metallicity higher than
solar are confirmed
by spectroscopic means (see Randich \cite{R07} and references therein).
Among them we mention
\object{NGC 6475} and \object{Hyades}
(ages $\sim$250 and 600 Myr, [Fe/H]=+0.14 and +0.13, respectively:
see Sestito et al.~\cite{sestito03} and Boesgaard \& Friel \cite{bf90}),
\object{Praesepe} ($\sim$600 Myr, [Fe/H]=+0.25; Pace et al. 2007, in preparation), the $\sim$1 Gyr old \object{NGC 6134} 
([Fe/H]=+0.15, Carretta et al.~\cite{carretta04})
and the 2 Gyr old \object{IC 4651} ([Fe/H]=+0.10: Pasquini
et al.~\cite{P04}; Carretta et al.~\cite{carretta04}).
The most noticeable
metal rich cluster
is the very old ($\sim$ 8--9 Gyr) \object{NGC 6791}, recently investigated by
Origlia et al.~(\cite{origlia}),
Carraro et al.~(\cite{carraro06}), and Gratton et al.~(\cite{G06})
who derive a metallicity [Fe/H]=+0.35, +0.39 and +0.47, respectively
(from giant stars).

The origin of metal rich disk clusters is puzzling,
especially in the case of old ones, 
since the classical view of Galactic evolution
predicts an over-time enrichment of the interstellar medium,
and as a consequence only the youngest stars should have metallicities
higher than solar.
Nevertheless,
other old and metal rich stellar 
populations exist, such as bulge field stars (Fulbright,
McWilliam, \& Rich 
\cite{fulb}, \cite{fulb07}) and objects in the solar neighborhood
with kinematics
and metallicities more similar to those of the bulge, in spite
of their position (Castro et al. ~\cite{castro};
Pompeia et al.~\cite{pompeia}).
The high metallicity of stars in the center of the Galaxy can be
explained with an early enrichment of that 
region, while the presence of old
metal rich stars/open clusters in the disk is more puzzling.
One hypothesis about the
origin of metal rich open clusters
is that they were born
in the inner side of the Galaxy, close to the bulge, where
the metal enrichment occurred early and rapidly,
and then they moved outwards in the disk. Alternatively, they 
might have been born in an external environment and then captured by our
Galaxy, as discussed
by Carraro et al.~(\cite{carraro06}) for NGC~6791, although a very recent paper
on this cluster excludes this possibility (Bedin et al.~\cite{bedin}).
Finally, the simplest explanation is that
metal rich clusters originated in the disk itself, in a region
characterized by faster enrichment. 
The element abundance distribution of metal rich
open clusters is very useful to put constraints
on their origin. 

Independently
on their origin, the very existence of metal rich open clusters
provides ideal samples to investigate
other topics, such as 
the dependence of light element depletion on chemical composition,
or planet formation and evolution.
In particular, it is now  well ascertained that stars hosting giant planets
are more metal rich than stars not harbouring planetary systems
(e.g. Santos~\cite{santos} and references therein). Metal rich
clusters thus represent very good targets where to search for 
planetary systems,
although photometric searches for transiting planets have 
not been successful so far (but have shown feasibility -- see, e.g.,
Paulson et al.~\cite{paulson1},
Paulson, Cochran, \& Hatzes \cite{paulson2} for the Hyades;
Mochejska
et al.~\cite{moche} for NGC 6791).

We present here a new high resolution spectroscopic investigation of 
NGC~6253; with respect to Carretta et al.~(\cite{carretta00}; 2007, submitted)
our sample covers a much wider region in
the color-magnitude diagram (CMD), including
not only a red clump member, but also 
turn off (TO) and subgiant/red giant branch (SGB, RGB) stars.
Since, as we will show in the paper, at very high
metallicities the Fe content of clump stars derived with equivalent
width
analysis
might be overestimated -- due to heavy line blending -- the analysis
of hotter TO and RGB stars should in principle provide more reliable results.
The paper is organized as follows: in Sect.~2 we describe the sample
and data reduction, while Sect.~3 is dedicated to the method of
analysis and estimate of uncertainties.
In Sect.~4 we report our results, checking the validity of the metallicity scale. The results are then discussed
in Sect.~5, and summarized in Sect.~6.

\section{Observations and data reduction}\label{obs}

The spectra of the NGC~6253 sample presented in this
paper were collected with FLAMES on VLT/UT2 
(Pasquini et al.~\cite{P00}), using the fiber link to UVES with 
a spectral resolution of $R=47,000$. 
The GIRAFFE fibers were instead used for collecting spectra of
a large number of main sequence stars, with the goal of
investigating lithium abundances and radial velocities (Randich et al. 2007,
in preparation).
The observations were carried out 
in service mode
during April 2004, with two FLAMES configurations
and using for each of them two different 
gratings (CD3 and CD4, covering the wavelength ranges
$\sim$4750--6800 \AA~and 6600--10600 \AA, respectively).
We used two configurations in order to maximize the
number of objects observed with GIRAFFE; as a consequence, also
the UVES pointings changed. More in detail, 
the two configurations differ for the
number of stars observed (seven and six stars were observed with UVES in configurations 
A and B, respectively) and for the correspondence between fiber and object.
Since all the six stars observed with UVES
in configuration B are in common with configuration A, we observed seven stars in total.
Table~\ref{obs6253} gives the log of observations for the cluster.
Data reduction was carried out by ESO personnel using the
dedicated pipeline, and we analyzed the 1-d, wavelength calibrated spectra
using standard  IRAF\footnote{IRAF is distributed by the National Optical
Astronomical Observatories, which  are operated by the Association of
Universities for Research in Astronomy,  under contract with the National
Science Foundation.}  packages. 
The contamination by atmospheric telluric lines was taken into account
by performing a correction on the spectra with the task TELLURIC in IRAF
via a comparison with early-type stars observed with UVES during another 
run. Background subtraction was carried out, as customary,
using one fiber dedicated to the sky.

\begin{table*}[!] \footnotesize
\caption{Observation log of NGC~6253.}\label{obs6253}
\begin{tabular}{llllll}
\hline
\hline
Date   & UT$_{\rm{beginning}}$ &Exposure time&Configuration &Grating& no. of stars\\
& &(s) & && \\
\hline
2004-04-07&07 31 40.549&2595 &A &CD3 & 7\\
2004-04-08&07 39 50.833&2595 &B &CD3 & 6\\
2004-04-06&08 34 34.455&2595 &A &CD4 & 7\\
2004-04-07& 08 27 40.831&2595&A &CD4 & 7\\
2004-04-08&08 37 19.858&2595 &B &CD4 & 6\\
\hline
\hline
\end{tabular}
\end{table*}

We report in Table~\ref{data6253} the target stars,
adopting
the ID numbers from the EIS survey (Momany et al. \cite{momany}; Col.~1);
since we used the photometry
by 
Bragaglia et al.~(\cite{bragaglia97}; Cols.~6 and 7),
we list for completeness also their IDs in Col.~2.
The only star not included in the study by Bragaglia et al.
is 105495, for which we adopted the EIS
photometry calibrated
to Bragaglia et al. As mentioned in Sect.~1, NGC 6253 was investigated
also by Twarog et al.~(\cite{ttl03}), therefore 
we provide a cross-identification
with their numbering system (Col.~3).
The signal-to-noise ($S/N$) ratios reported in Col.~10
have been measured in the spectral regions
around 5600 \AA~and 6300 \AA.

\begin{table*}[!] \footnotesize
\caption{Data for NGC 6253. ID$_{\rm{B97}}$ and
$BV$ photometry (non corrected for reddening)
are from Bragaglia et al.~(\cite{bragaglia97}) except for
star 105495, see text;
we report also ID$_{\rm{EIS}}$, used through all the paper,
and a cross-identification with Twarog et al.~(\cite{ttl03}).
The number of exposures
for each star is intended as number of pointings
with the same cross-disperser. }\label{data6253}
\begin{tabular}{llllllllllllll}
\hline
\hline
Star & Star  &   Star & RA	     &     DEC      & $V$      &  $B$     & no. exp. & $RV\pm{rms}$   & $S/N$&Notes \\
ID$_{\rm{EIS}}$&ID$_{\rm{B97}}$ &ID$_{\rm{T03}}$  & & & & &CD3/CD4 &(km s$^{-1}$)&& \\
\hline
069885 & 3053 &175& 16 58 52.311& $-$52 41 40.76 & 14.400 & 15.210 &2/3 & $-$28.81$\pm$1.30 &70--80 & TO, M     \\
023501 & 2225 &160& 16 58 53.250& $-$52 42 43.32 & 14.330 & 15.192 &2/3 & $-$33.49$\pm$2.87 &65--90 & TO, NM? \\
069360 & 2542 & 73 & 16 59 07.044& $-$52 42 21.84 & 13.229  & 14.520&1/2 & $-$21.07 $\pm$0.59&100--120 & RGB, NM\\
022182 & 1556 &127  &16 59 15.518& $-$52 43 33.11 & 13.865  & 15.094 &2/3 & $-$35.51$\pm$0.77&90--110 & SGB/RGB, NM\\
023498 & 2253 &133& 16 59 21.358& $-$52 42 43.18 & 13.933  & 14.886&2/3 & $-$30.13$\pm$1.01 &65--100 & SGB, M\\
024707 & 3138 & 97 & 16 59 12.924& $-$52 41 35.76 & 13.548  & 14.726&2/3 & $-$30.58$\pm$0.65 &75--100 & SGB/RGB, M\\
105495 & --    &52& 16 58 51.192& $-$52 36 57.47 & 12.700  & 14.110&2/3 & $-$29.33$\pm$0.84 &120--140 & clump, M \\
\hline
\hline
\end{tabular}
\end{table*}

\begin{figure*}
\psfig{figure=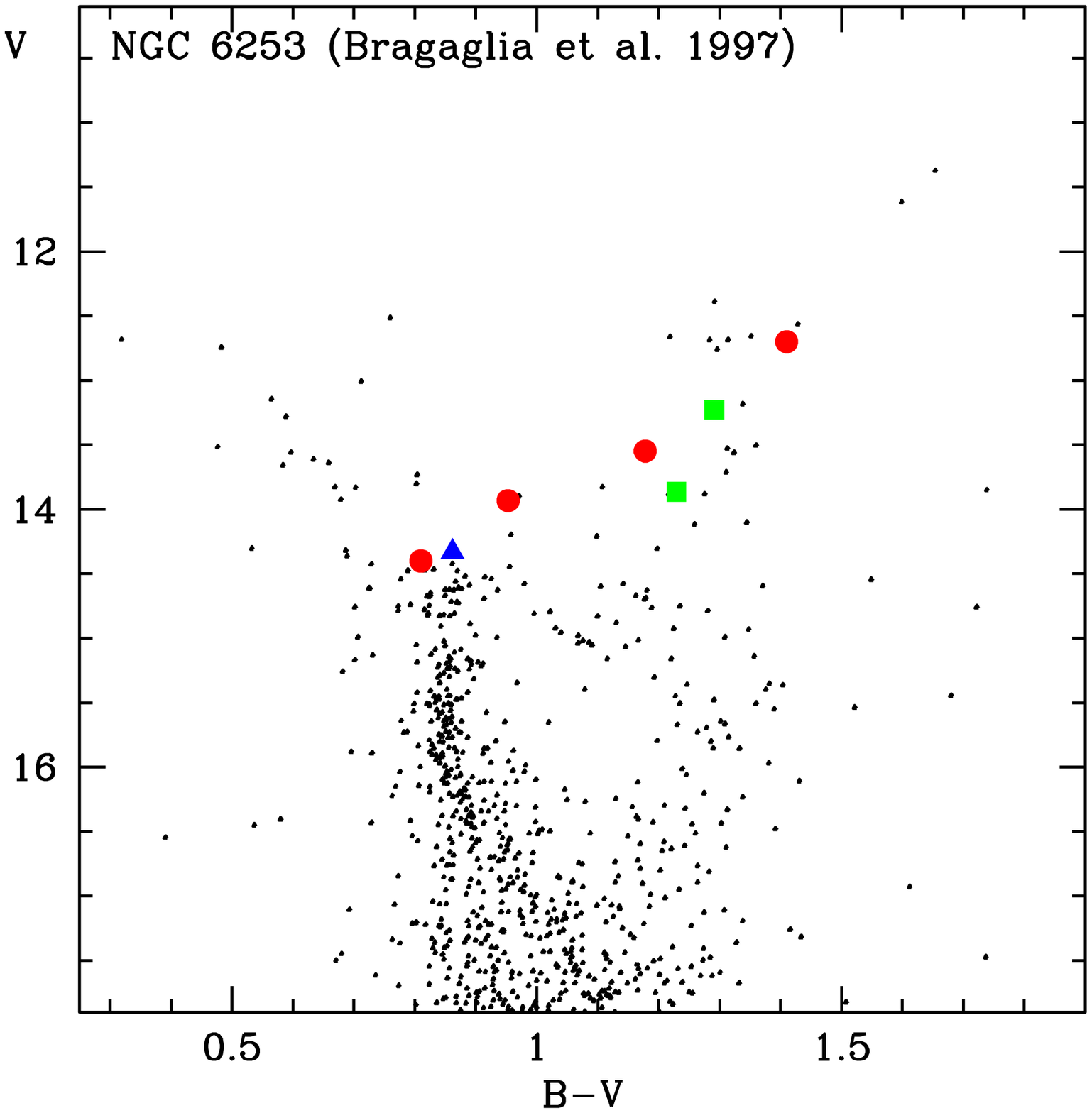, width=10cm, angle=0}
\caption{Color-magnitude diagram for NGC~6253. 
The observed stars are evidenced
by circles (members), triangles
(the radial velocity doubtful member, but
with metallicity consistent with membership) and squares (non-members).}\label{CMD}
\end{figure*}

We measured radial velocities ($RV$) with RVIDLINES using several tens of metallic
lines on each single spectrum, and subsequently we combined
multiple spectra. The heliocentric $RV$s (Col.~9 of Table 2) have 
uncertainties  of $\sim$1 km s$^{-1}$
with the exception of star 023501 for which the error is almost 
3 km s$^{-1}$.
The mean $RV$ of the whole sample is $-29.85\pm$4.55 km s$^{-1}$: $RV$s
of stars
069360 and 022182 deviate from this value
by more than 1$\sigma$; therefore we consider them non-members,
although we cannot exclude that they are binary cluster members.
If we exclude these two stars, we obtain $<RV>=-30.47\pm1.82$ km s$^{-1}$
for the remaining 5 objects.
However, also star 023501 has a $RV$ slightly deviating from the
two averages above (by $\sim$3 km s$^{-1}$), therefore we will provisionally consider it
as a doubtful member (see below and Sect.~\ref{results}).
By computing the average radial velocity considering only the 
4 stars which can be safely classified as members,
we have $<RV>=-29.71\pm0.79$ km s$^{-1}$.
Figure~\ref{CMD} shows the CMD of
the cluster, where the selected stars are marked
using different symbols: circles (members),
squares (non-members) and triangle for star 023501.
Note that
the positions in the CMD of
stars with radial velocity deviating from the mean
are consistent with membership (in particular
that of the doubtful member 023501).
Notes on the evolutionary status and membership of the stars are shown in
Col.~11 of Table~\ref{data6253}.

\begin{figure*}
\psfig{figure=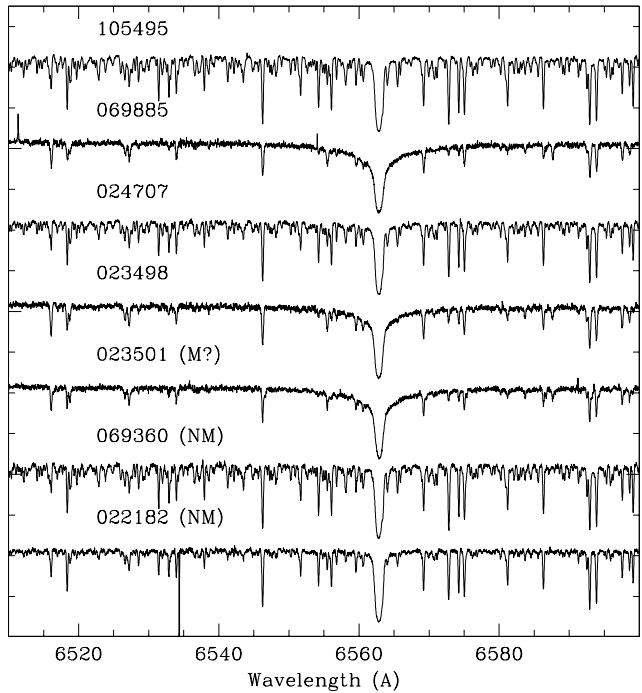, width=10cm, angle=0}
\caption{NGC~6253 sample spectra in the wavelength region
at 6500--6600 \AA.}\label{spettri}
\end{figure*}

Finally, in Fig.~\ref{spettri} we show the spectra
(referred to $RV$=0) in the wavelength region around the H$\alpha$ feature
for all the observed stars.

\section{Analysis}\label{moog}

\subsection{Line lists and equivalent widths}\label{linelist}

The analysis of chemical abundances was performed
by means of equivalent widths ($EW$s) using
an updated version (2006) of the package MOOG (Sneden
\cite{sneden})\footnote{http://verdi.as.utexas.edu/} and using model
atmospheres by Kurucz (\cite{kuru}). MOOG works under the assumption
of local thermodynamic equilibrium (LTE).

Solar abundances of Fe and other elements (Na, Mg, Si, Ti,
Cr, Ni, Ba) were derived in order to determine the
zero point of the metallicity scale. 
The line lists adopted for the Sun and for evolved stars
were retrieved from
Gratton et al.~(\cite{G03}) and
are described in
Paper~{\sc i} and 
Bragaglia et al.~(2007, in preparation).
We recall from Paper~{\sc i} 
that when available we adopted
collisional damping coefficients from Barklem, 
Piskunov \& O'Mara (\cite{barklem}),
otherwise we considered
the coefficients by Gratton et al.~(\cite{G03}), or,
for a few lines, the classical Uns\"old (\cite{unsold}) approximation.
For the Sun we obtain  $\log n(\rm{Fe}$~{\sc
i})=7.49$\pm$0.04 (standard deviation, or rms) using
\teff=5779 K, $\log g$=4.44, and $\xi$=0.8 km
s$^{-1}$.

The spectra were normalized using CONTINUUM in IRAF,
dividing the spectra in small regions (50 \AA)
and visually checking the output; $EW$s for the various lines were
measured with SPECTRE (developed and maintained by C.~Sneden)
by  Gaussian fitting of the line profiles.
We provide the $EW$s for each star in Table 3 (available only in
electronic form):
the first two Cols.~list the wavelength and the element,
and the others show 
the corresponding $EW$ for each star. 
Note that the definition of the continuum and the determination
of the $EW$s are a very critical step for our stars, since they are
rather cool objects with low gravity,
and with an exceptionally high metallicity.
As a consequence, several lines can be affected by strong blending
(see also below).

\subsection{Stellar parameters}\label{para}

Initial effective temperatures (\teff) and gravities were 
estimated from 
photometry. In the case of giant stars we used the 
$B-V$ vs.~\teff~calibration by Alonso, Arribas, \& Martinez-Roger
(\cite{alonso99}) for giants, while
for TO stars we adopted the calibration by 
Alonso, Arribas, \& Martinez-Roger (\cite{alonso96}), 
based on a large sample of dwarfs.
Surface gravities were
derived using the
expression $\log g=\log (M/M_{\odot})+0.4(M_{bol}-M_{bol{\odot}})+4{\cdot}\log (T_{\rm {eff}}/T_{\rm{eff}\odot})+\log g_{\odot}$, where $M$ is the mass and $M_{bol}$ 
the bolometric magnitude (with 
$M_{bol{\odot}}$=4.72). Bolometric corrections for giants
were derived following Alonso et al.~(\cite{alonso99}), while
those for the two TO stars were retrieved from Johnson (\cite{john66})
and are close to 0.
We adopted the most recent cluster parameters by 
Bragaglia \& Tosi (\cite{bt06}), 
$(m-M)_{0}=11.0$, $E(B-V)=0.23$, and an age of 3 Gyr, which,
using $Z=0.05$ isochrones, corresponds to masses $M\sim$1.32 $M_{\odot}$
at the TO and $\sim$1.40 $M_{\odot}$ at the clump. We assumed $M=1.40\,M_{\odot}$
for all the stars, since uncertainties $\lesssim$ 0.1 $M_{\odot}$
translate into differences of 0.03 dex or lower in surface gravities, which
are well below the random errors and do not affect the metallicities
derived from Fe~{\sc i} lines.
Note that the calibrations by Alonso et al.~are valid up to a metallicity
of [Fe/H]=+0.2, while the cluster should have a higher Fe content; 
nevertheless,
the weak dependence of the \teff~on $B-V$ colors suggests us that the error
committed is small, and in any case the photometric temperatures are
used only
as initial parameters and are optimized during the spectroscopic analysis.

In Paper~{\sc i} 
the microturbulent velocities were derived
using the relationship by Carretta et al.~(\cite{carretta04}), 
based on the
optimization of Fe abundance as a function of \emph{theoretically expected} 
$EW$s for the given
lines (see the quoted references for further details).
The formula cannot be safely applied for NGC~6253, since it was 
based
on a sample of giants in clusters with 
metallicities closer to solar; in this case we deal instead with a very metal-rich cluster and also
with TO stars. Therefore, we used the microturbulences
by Carretta et al.~only as starting values; then, we optimized them
by minimizing the slope of
$\log n$(Fe)~{\sc i} vs.~the \emph{observed} $EW$s; when
plotting Fe abundances as a function of \emph{expected} $EW$s, we obtain
a rather small trend (i.e. the slope of
the relationship is smaller than its error), suggesting that the two methods of analysis
are in fair agreement. 

As customary, final effective temperatures were derived during the analysis
(after 1$\sigma$ clipping) by minimizing the trend
of $\log n$(Fe~{\sc i}) vs.~the excitation potential ($EP$);
as for surface gravities, 
we cannot derive them from
the ionization equilibrium condition (i.e. the assumption that
the difference between Fe {\sc i} and Fe {\sc ii} abundances
in the stars analyzed
should be similar to that found for the Sun)
since the few lines of
ionized Fe are strongly affected by blending.
On the other hand, since in most cases
spectroscopic temperatures are in good
agreement with photometric ones and the distance
is well known, we decided to adopt the
photometric $\log g$ values and left them unchanged during the analysis.
The only stars for which the spectroscopic and photometric \teff~differ by
a rather significant amount
($\sim$200 K) are the TO stars;
note however that changing the \teff~by $\pm$200 K
would imply a variation of $\sim\pm$0.1 dex in $\log g$, consistent with
the errors (see Sect.~\ref{errori}).
For the reasons mentioned above, we retained in the analysis only
neutral Fe features.

In Table~\ref{tab_Fe} we list
for each star the photometric
and spectroscopic \teff~(Cols.~2 and 4),
the adopted surface gravity (photometric, Col.~3), and
the spectroscopic microturbulence (Col.~5).

\subsection{Errors}\label{errori}

The major source of random
uncertainties affecting element abundances
derive from errors in $EW$s and uncertainties
in stellar parameters; systematic uncertainties
come from biases due to the method of analysis
adopted
and from errors in the line list (i.e. possible blendings and
oscillator strengths).

The errors in abundances related to $EW$s are given in good approximation
by the standard deviation (rms) around the mean abundance derived 
from individual lines for each star, call it $\sigma_1$.
This rms includes also the errors related to atomic parameters:
$\log gf$ values were taken from the literature, then 
we cannot give a precise estimate of the effect of their uncertainties; 
however, since
our abundance scale is directly referred to solar abundances
and the line list used for cluster stars is 
very similar to that adopted for the Sun, we can assume that
internal errors due to uncertainties in the atomic parameters
are minimized.
Note that when the abundances of elements other than Fe
are expressed as [X/Fe]\footnote{[X/Fe]=$\log n$(X)$-\log n$(X)$_{\odot}$$-$[Fe/H]
for each star, or
[X/Fe]=[X/H]$-$[Fe/H].}, a total $\sigma_1$
should be computed by quadratically adding the rms $\sigma_1$
on [X/H] and on [Fe/H].

The contribution of uncertainties in stellar parameters,
\teff, $\log g$ and $\xi$,
were estimated by varying each parameter of a given quantity (leaving
the other two unchanged) and then adding the three errors.
We assumed variations of $\pm$70 K in \teff~and $\pm$0.10 km s$^{-1}$
in $\xi$, since these changes would introduce significant trends into
the relationship of Fe abundances with $EP$ and observed $EW$s.
We cannot estimate in a similar way an uncertainty in $\log g$,
since we did not
optimize the gravity using the ionization equilibrium.
However, usually errors in $\log g$ are of the order 
of $\sim$0.15--0.25 dex (e.g., Paper~{\sc i});
thus, we assumed a conservative $\Delta{\log g}=\pm0.25$ dex.
Table~\ref{sens} shows the sensitivity of elemental abundance
($\log n$(X))
to variations in the atmospheric parameters for two cluster members:
the TO star 069885 and the clump star 105495.
In the case of Fe we computed $\sigma_2$, 
the quadratic sum of the three errors due to uncertainties in the 
stellar parameters; this was not done for the other elements, since
in the final computation of [X/Fe]
one should take into account the $\sigma_2$ for [X/H] and for [Fe/H], which
could go into opposite directions (see Table~\ref{sens}).

Finally, we wish to give an estimate of the systematic uncertainties 
in the Fe abundance scale related to
the method of analysis. This can be done for example by
analyzing a star with a well known metallicity, possibly observed 
with the same instrument and
using the same method
of analysis.
Since we did not collect spectra of stars outside of the clusters
included in the program,
we performed the analysis for two clump stars in the Hyades
observed with SARG at TNG at similar resolution.
A detailed description of the analysis of the two Hyades
is reported in Paper~{\sc i}; here, we only mention that we did find
a [Fe/H] in reasonable 
agreement with the literature estimates, confirming that,
up to the metallicity of the Hyades,
our method of analysis should not be affected by large systematic errors.
In order to check if the metallicity scale is correct also for
a very  high metal content, we carried out some tests on 
$\mu$ Leonis (see Sect~\ref{muleo}), a rather
luminous giant which is known to have a
remarkably oversolar metallicity (e.g. Gratton \& Sneden \cite{gs90}).
We anticipate here that at the metallicity of NGC~6253 our Fe
abundance for the coolest star
(clump) might be overestimated by $\sim$0.1--0.15 dex.

\section{Results}\label{results}

\setcounter{table}{3}
\begin{table*}[!] \footnotesize
\caption{Stellar parameters and Fe abundances
for stars in NGC~6253. Numbers in parenthesis in Col.~6 are
[Fe/H] estimated by spectral synthesis.}\label{tab_Fe}
\begin{tabular}{llllllllllllll}
\hline
\hline
Star & \teff$_{\rm{phot}}$ & $\log g_{\rm{phot}}$ & \teff$_{\rm{spec}}$ & $\xi$ & [Fe/H] &$\sigma_{1}$ & $\sigma_{\rm{tot}}$ & Membership \\
 & (K) & & (K) & (km s$^{-1}$) & & &\\
\hline
069885 & 6032 & 3.84 & 6200 & 1.27 & +0.45(+0.35) &0.09 & 0.11& M\\
023501 & 5842 & 3.75 & 6050 & 0.90 & +0.29 &0.09 & 0.11& M?\\
023498 & 5509 & 3.44 & 5630 & 1.30 & +0.32 &0.09 & 0.11& M\\
024707 & 4972 & 3.05 & 4940 & 1.16 & +0.39 &0.09 & 0.11& M\\
105495* & 4520 & 2.46 & 4450 & 1.23 & +0.49(+0.35) &0.10 & 0.13& M\\
Average Fe& & & & &+0.39(+0.34)&& 0.08(0.04) (rms)& \\
\hline
Likely non-members \\
069360 & 4741 & 2.81 & 4680 & 1.20 & +0.48 &0.10 & 0.12& NM\\
022182 & 4865 & 3.13 & 4770 & 1.20 & +0.12 &0.07 & 0.09& NM\\
\hline
\hline
\end{tabular}
* Note:  the initial parameters of the clump star have been computed from EIS magnitudes
calibrated to Bragaglia et al.~(1997, see Table 2). 
We also performed a test using the Str\"omgren photometry
by Twarog et al.~(2003) reported to the $UBV$ system. 
In this case we found \teff$_{\rm{phot}}$=4666 K and
$\log g$=2.58, and, from the spectroscopic analysis, \teff$_{\rm{spec}}$=4500 K, $\xi$=1.23 km s$^{-1}$, 
[Fe/H]=+0.51 ($\log g$=2.58), which agree within the errors with the values reported in Table \ref{tab_Fe}.
\end{table*}

\begin{table*}[!] \footnotesize
\caption{Sensitivities of abundances
($\log n$(X)) to variations in the atmospheric parameters for 
TO and clump stars in NGC~6253.}\label{sens}
\begin{tabular}{ccccccccccccccc}
\hline
\hline
$\Delta$ &$\sigma_{T\rm{eff}}$ &$\sigma_{\log g}$ &$\sigma_{\xi}$  \\
&$\Delta$\teff=$\pm$70 K & $\Delta{\log g}$=$\pm$0.25 dex & $\Delta{\xi}$=$\pm$0.10 km s$^{-1}$\\
\hline 
069885 & & & \\
Fe~{\sc i}&$+0.05/-0.05$ &$-0.02/+0.03$ &$-0.03/+0.03$ \\
Na~{\sc i}&$+0.03/-0.04$ &$-0.05/+0.04$ &$-0.02/+0.01$ \\
Mg~{\sc i}&$+0.03/-0.03$ &$-0.03/+0.03$ &$-0.02/+0.02$ \\
Si~{\sc i}&$+0.03/-0.02$ &$-0.01/+0.02$ &$-0.01/+0.02$ \\
Ca~{\sc i}&$+0.05/-0.04$ &$-0.05/+0.05$ &$-0.03/+0.04$ \\
Ti~{\sc i}&$+0.06/-0.07$ &$-0.01/+0.01$ &$-0.03/+0.02$ \\
Cr~{\sc i}&$+0.05/-0.05$ &$-0.01/+0.02$ &$-0.03/+0.03$ \\
Ni~{\sc i}&$+0.05/-0.04$ &$0.0/+0.01$ &$-0.04/+0.03$ \\
Ba~{\sc ii}&$+0.02/-0.02$ &$+0.04/-0.05$ &$-0.07/+0.07$ \\
105495 & & & \\
Fe~{\sc i}&$+0.01/+0.03$ &$+0.08/-0.01$ & $-0.03/+0.07$\\
Na~{\sc i}&$+0.07/-0.04$ &$-0.05/+0.04$ &$-0.02/+0.05$ \\
Mg~{\sc i}&$0.00/+0.01$ & $+0.02/+0.01$& $-0.03/+0.04$ \\
Si~{\sc i}&$-0.05/+0.06$& $+0.07/-0.02$&  $-0.02/+0.03$\\
Ca~{\sc i}&$+0.08/-0.05$& $-0.03/+0.06$&  $-0.04/+0.07$\\
Ti~{\sc i}&$+0.12/-0.06$& $+0.03/+0.02$&  $-0.05/+0.11$\\
Cr~{\sc i}&$+0.08/+0.02$ & $+0.03/+0.02$&  $-0.05/+0.09$\\
Ni~{\sc i}&$+0.01/+0.04$ & $+0.09/-0.02$&  $-0.04/+0.08$\\
Ba~{\sc ii}&$+0.02/-0.02$& $+0.05/-0.05$&  $-0.06/+0.07$\\
\hline
\end{tabular}
\end{table*}

\subsection{Metallicity}\label{ferro}

Fe abundances are listed in
Table~\ref{tab_Fe}, together with their errors, in Cols.~6--8:
$\sigma_1$, the standard deviation
from the mean abundance obtained over the whole set of lines for each star,
and $\sigma_{\rm{tot}}$, the
total uncertainty in which we consider also errors
due to stellar parameters $\sigma_2$  
($\sigma_{\rm{tot}}=\sqrt{{\sigma_1}^{2} + {\sigma_2}^{2}}$).
Star 023501 was classified as a doubtful member
from its radial velocity (see Sect.~\ref{obs}): we found for it
[Fe/H]=+0.29 in agreement with those of the confirmed members, thus we
conclude that this object is a probable cluster member.
Stars 069360
and 022182 are instead radial velocity non-members  but, as
  already mentioned, we cannot exclude that they are binary cluster
  members.
The first one
has a high Fe abundance (+0.48)
similar to those of members; nevertheless,
since we do not optimize gravities from the ionization equilibrium
 and we rely on the photometric
values, the derived [Fe/H] value might be due only to
a coincidence; in other words, if the adopted distance is 
wrong, one finds a wrong metallicity. On the other hand, if the
  star
would effectively be a binary cluster member, the [Fe/H] found by us
might be the correct one.
Star 022182
has a much lower [Fe/H] with respect to other stars,
that is +0.12.
In any case, we disregard the two non-members in the following.
The average metallicity (computed excluding
the non-members) with the rms error
is also shown in Table~\ref{tab_Fe}, [Fe/H]=+0.39$\pm$0.08.


For the 4 ascertained members we find [Fe/H]=+0.45 (TO star),
+0.49 (clump) and +0.32, +0.39 (SGB/RGB).
The metallicity of the clump star 105495 based on $EW$s could likely
 be overestimated, due to unresolved blends (the spectrum is
very crowded due to the combination
of high Fe content and low temperature);
by excluding the clump star, the average
metallicity
slightly decreases to [Fe/H]=+0.36$\pm$0.07.
A possible offset in the metallicity scale will 
be discussed in the next sections (\ref{sintesi} and \ref{muleo}).
Values shown in parenthesis in Col.~6
of Table~\ref{tab_Fe} are the metallicities found from spectral synthesis
for the hottest TO star and for the clump star;
the average (+0.34) computed taking into account these values is also reported.

\begin{figure*}
\psfig{figure=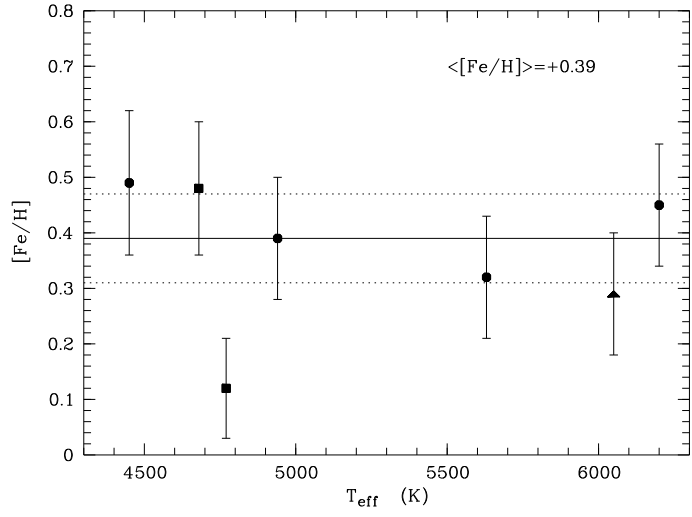, width=7cm, angle=-90}
\caption{Fe abundances as a function of \teff~for stars in NGC~6253.
Symbols for members, the possible member
and non-members are the same as in Fig.~\ref{CMD}.
The solid line represents the average Fe content (including all the
stars but the
non-members), while the dashed ones
represent the rms.}\label{Fe_Teff}
\end{figure*}

Figure~\ref{Fe_Teff} shows [Fe/H] values as a function of 
effective temperature for all the stars observed. 
The solid and dotted lines indicate
the average cluster metallicity $\pm$ rms ([Fe/H]=+0.39$\pm$0.08). 


\subsection{Spectral synthesis}\label{sintesi}

In order to check Fe abundances derived through the $EW$ analysis, we
carried out spectral synthesis for the warmest (069885) and
coolest (105495) cluster stars,
which are also those having the highest [Fe/H].
The spectral synthesis was performed in
a spectral interval of $\sim \pm 10$~\AA~around the Li~{\sc i} 6707.8~\AA~line.
As for the $EW$ analysis, we used MOOG and Kurucz atmospheres; an earlier
version of MOOG (2000) was however employed, with
a line list optimized to fit the solar spectrum obtained
with UVES; the Fe abundance retrieved from the synthesis are therefore
differential with respect to the Sun.
We used
the classical 
Uns\"old (\cite{unsold}) approximation for the damping coefficients,
since the spectral range investigated with the synthesis does not
include strong lines.

\begin{figure*}
\psfig{figure=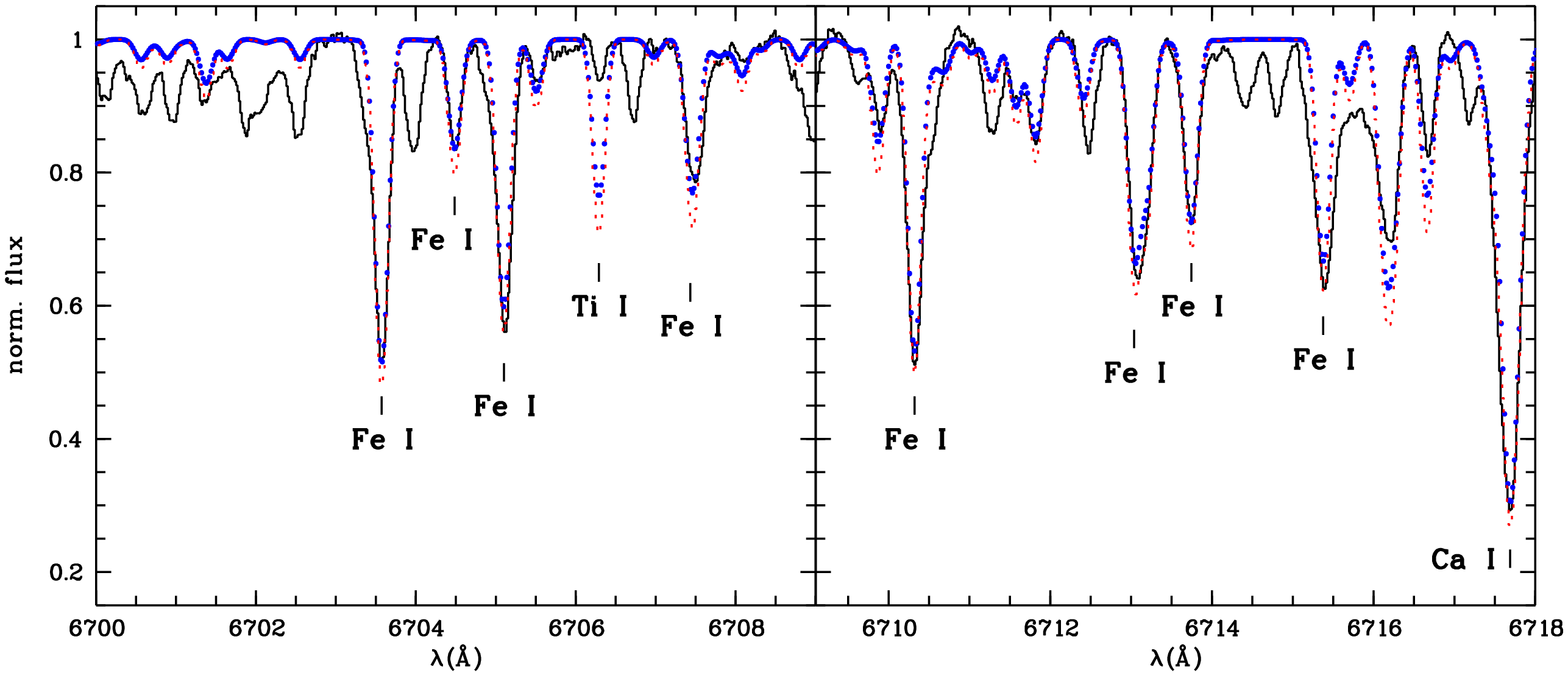, width=10cm, angle=-90}

\vspace{-1cm}

\psfig{figure=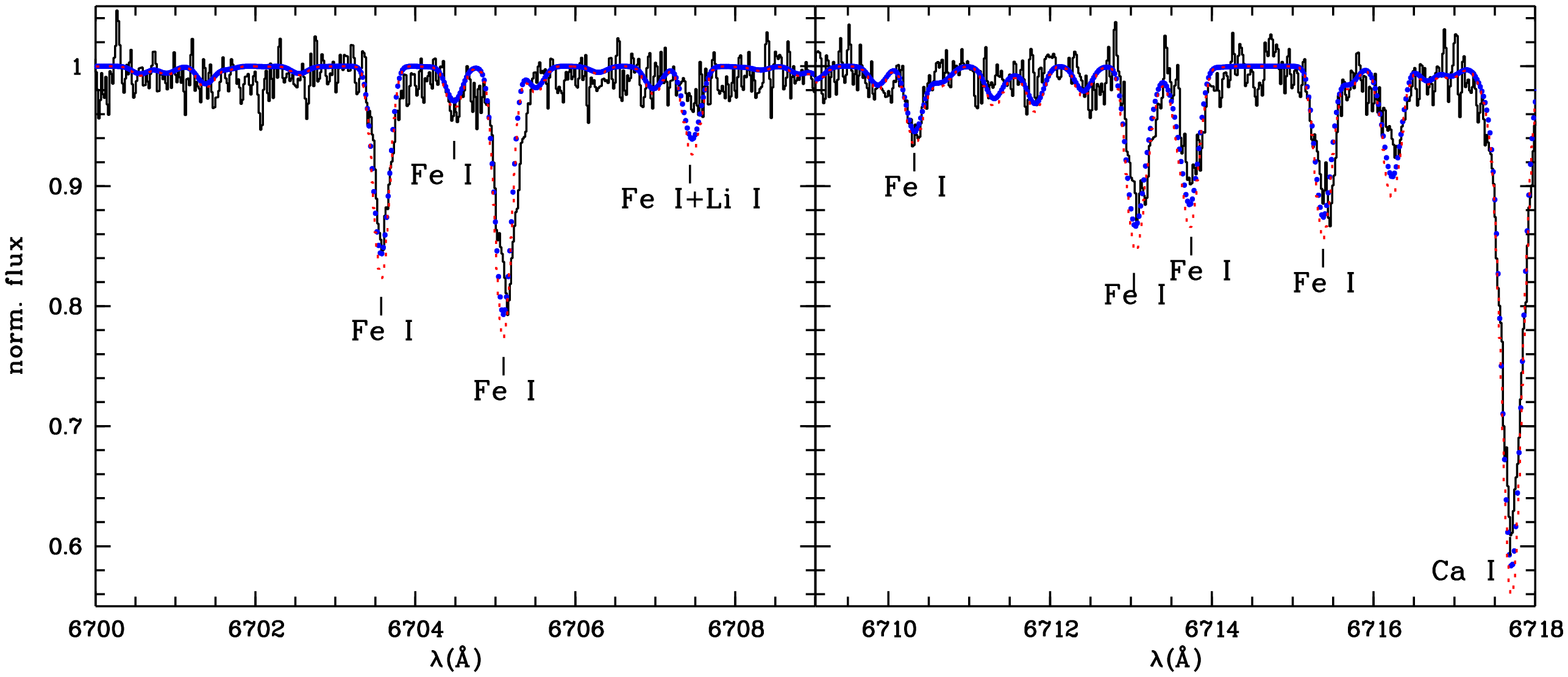, width=10cm, angle=-90}
\caption{Comparison between observed (solid line) and synthetic spectra
for the clump star 105495 (upper panel) and the TO star 069885 (lower panel) in the Li~{\sc i} region (6707.8 \AA).
The (red) thin dots are spectra computed adopting the spectroscopic
metallicities (+0.49 for 105495
and +0.45 for 069885), while the (blue) thick dots are spectra with
[Fe/H]=+0.35.}\label{synt}
\end{figure*}

Synthetic spectra were computed adopting stellar parameters derived
by the $EW$ analysis and listed in Table~\ref{tab_Fe}. 
For each star, we computed
a synthetic spectrum with the metallicities determined through $EW$s and
then others until the best fit of the observed spectrum was obtained.
Figure~\ref{synt} shows the spectral synthesis
for two stars (105495 and  069885) in the wavelength range
6700--6718 \AA.
In the upper panel, we plot the comparison between the observed spectrum of 
the clump star (solid black line)
and two synthetic spectra with [Fe/H]=+0.35 (blue thick dots) and +0.49
(red thin dots):
as evident, the lines of Fe~{\sc i}
in the synthetic spectrum with metallicity +0.49 are in the majority of cases deeper
than the observed ones, while the synthesis
with [Fe/H]=+0.35 provides a better fit (although far from perfect).
This happens also for the TO star (lower panel)
for which we computed two synthesis with [Fe/H]=+0.35 and +0.45 (the symbols
are the same as in the upper panel).
Therefore, in both cases the metallicity obtained from
spectral synthesis is  $\sim 0.1$~dex
below that determined using $EW$s.
A higher metallicity from $EW$ analysis
with respect to synthesis 
can be easily explained for
the clump star, whose spectrum 
might be affected by blending; on the other hand,
the discrepancy found for the TO star is rather 
surprising, and we do not really have an explanation for this.

As mentioned in Sect.~\ref{ferro}, by assuming [Fe/H]=+0.35
for the TO and clump stars, the average [Fe/H] of the cluster would 
decrease down to +0.34. 
However, we stress 
that errors in the determination of the continuum in the observed spectra
affect also the comparison with synthetic spectra and not only
$EW$ measurements.
Finally, note that in the wavelength region shown in Fig.~\ref{synt}
there are observed spectral features not reproduced by the synthesis:
these are lines of elements other than Fe, which in this case are not important for
the determination of the metallicity (e.g. the
Ti~{\sc i} line at 6706.3 \AA~and the Li~{\sc i} line at 6707.8 \AA)
and for which, therefore, we did not optimize atomic parameters
and abundances.

\subsection{Comparison with $\mu$ Leo}\label{muleo}

Since the analysis through $EW$s and the spectral synthesis yield
slightly different results, we
carried out a further test
in order to check our metallicity scale: namely, we analyzed
the metal rich giant star $\mu$ Leonis.
The values of [Fe/H] estimated in the literature are all around +0.3--0.4 dex
(e.g., Gratton \& Sneden \cite{gs90}, [Fe/H]=+0.34; Fulbright,
\cite{fulb}, [Fe/H]=+0.32; Gratton et al. \cite{G06},
[Fe/H]=+0.38), therefore similar to that of NGC~6253.
 
\begin{figure*}
\psfig{figure=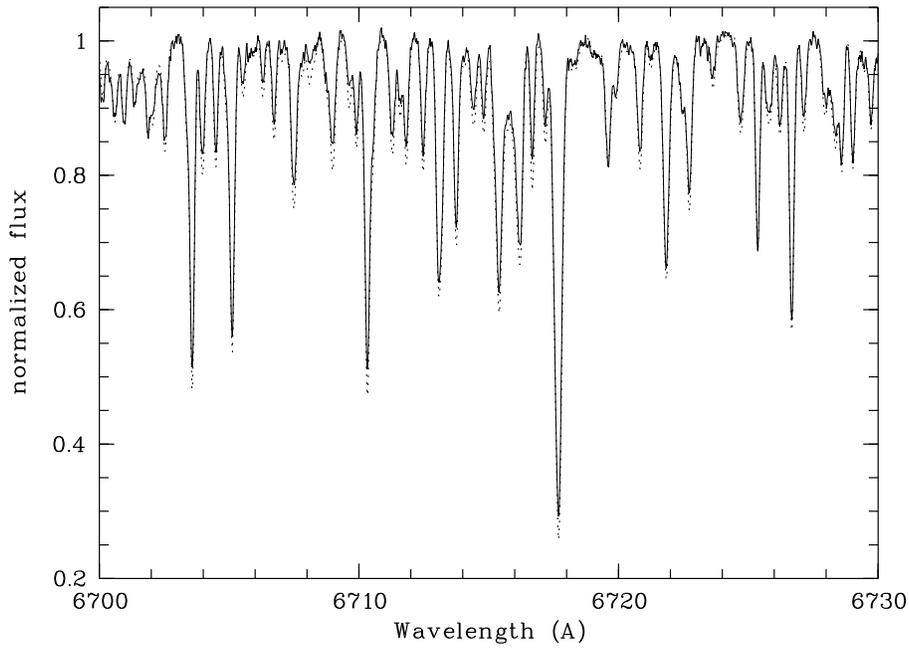, width=9cm, angle=-90}
\caption{Comparison between the spectra of the
clump star 105495 in NGC 6253
(solid line) and of $\mu$ Leo (dashed line) 
in the region of the Li~{\sc i} feature (6707.8 \AA).}\label{fig_muleo}
\end{figure*}

We analyzed a  spectrum of $\mu$ Leo
observed with FEROS on the 2.2m Telescope at La Silla Observatory
with a resolution similar to that of our sample stars.
By using the same line list as for NGC~6253, we obtained [Fe/H]=+0.51 for 
$\mu$ Leo, and \teff=4400 K, $\log g$=2.3 and $\xi$=1.2 km s$^{-1}$; whereas
the atmospheric parameters are in good agreement
with the determinations of other authors,
the Fe abundance is higher than previously found.
A metallicity more similar to those quoted by other authors is obtained by us
with the spectral synthesis, i.e. [Fe/H]=+0.38;
since by spectral synthesis we find [Fe/H]=+0.35 for 105495, which has atmospheric
parameters similar to those of $\mu$ Leo, our results
suggest that the clump star in NGC~6253 and $\mu$ Leo should actually
have similar metallicities, but a scale offset is present between the 
synthesis and $EW$ analysis.
Figure~\ref{fig_muleo} shows a comparison between 
the spectra of $\mu$ Leo (dashed line) and 105495 (solid line)
in the wavelength region around the Li~{\sc i} doublet at 6708 \AA,
where several Fe~{\sc i} features are present:
as clearly visible, the spectral lines of the two stars are similar,
 suggesting
that the metallicities are nearly the same. In particular,
$\mu$ Leo is slightly
colder than 105495, and indeed its metal lines
are slightly stronger.
In any case,
the problem of the determination of
a zero-point for the [Fe/H] scale remains.

In order to further check the metallicity
scale and the origin of the discrepancy, 
we carried out different tests on $\mu$ Leo.
We considered the line list for Fe~{\sc i} adopted by Fulbright 
et al.~(\cite{fulb}): more in detail, using the 30 lines in common
with our list we carried out an $EW$ analysis adopting our atomic 
parameters and their $EW$s.
The analysis by Fulbright et al.
is differential with respect to the Sun, for which
they derive $\log n$(Fe)$\odot$=7.45.
With their measurements and our atomic parameters we find [Fe/H]=+0.40,
i.e. 0.08 dex larger than that of Fulbright et al.
(+0.32). 
Note that  we obtain \teff=4550 K, higher
than previously found by us, but in agreement
with Fulbright et al.~(\cite{fulb}).
If we adopt the 30 lines in common with Fulbright et
al., but using our $EW$s, we find a metallicity [Fe/H]=+0.49,
with \teff=4450 K.
Finally, we repeated the latter analysis also for the clump star
105495 in NGC 6253, i.e. with the lines in common with
Fulbright (and obviously our $EW$s), and 
we obtained [Fe/H]=+0.44.
Part of the discrepancies with previous analysis can be due
to $EW$ measurements (program/method and continuum tracing)
and the adopted code, but in any case we obtain a systematically higher
metallicity. 

Given the disagreement found with the literature results
for $\mu$ Leo, and between the spectral synthesis and $EW$ analysis for
stars in NGC~6253,
we conclude that
an offset 
in the abundance of the clump star is present
(due probably to a combination of low \teff~and high [Fe/H]
which results into very strong and blended features); whereas
we are not able to precisely quantify the offset,
we note that it is in the range $\sim$+0.10--0.15 dex; 
for the other stars we cannot estimate a possible abundance shift
through a  similar comparison with a well known star,
since we do not know  a very metal rich subgiant to be used
as a reference object.
In summary, we will adopt for all the stars the [Fe/H] values derived
through $EW$ analysis, but with the \emph{caveat} that those of the clump
and of the hottest TO stars are probably overestimated,
while
the metallicity found for the remaining stars is likely to be correct.

As far as other elements are concerned, it has not been possible to
carry out a direct comparison with $\mu$ Leo.
Although a chemical analysis of this star has been
recently performed by Fulbright (\cite{fulb07}),
their line lists and ours have only a small number of
lines (if any) in common for each element; moreover, their
$EW$s were not published.
By comparing the abundances 
for $\alpha$-elements (Ca and Si in particular) obtained using 
the few lines in common with Fulbright and our $EW$s, we obtain [X/H] values
similar to theirs, suggesting that the scale offset is likely to
affect only Fe.

\subsection{Abundances of other elements}\label{altri}

Besides Fe, we derived the abundances of the light isotope
Na, the $\alpha$-elements Mg, Si, Ca, Ti,
the Fe-peak elements Cr and Ni and the s-process element Ba.
[X/H] values for Si, Ca, Ti, Cr, Ni and Ba
are listed in Table \ref{alphaXH}, together
with the errors due to uncertainties in $EW$s ($\sigma_1$).
The $\log n$(X) values in the Sun found by us (with the exception of Ba), are also shown (last Col.).
[X/Fe] ratios are instead shown in Table~\ref{alpha},
with errors
computed by quadratically adding the rms $\sigma_1$ for [Fe/H] and for [X/H].
We notice that for Ba and Si
the measurement is based on
a small sample of lines (3 and 4--6 features,
respectively). 
For the other species (Ca, Ti, Cr and Ni) we carried out the analysis
using a 
large set of lines, and we performed 1$\sigma$ clipping.
In the last three Cols.~of Table 
\ref{alpha} we report
two average values ($\pm$  rms) of the [X/Fe] abundances for the cluster.
The means
are computed including all the five stars (Average1),
or
excluding the clump star (Average2).
In general the two averages are consistent within the errors 
one with each other; the scatter is smaller if only the confirmed members are considered.

As discussed in further detail by Bragaglia et al. (2007, in preparation),
we find a rather large error ($\pm$0.14)
on the solar $\log n(\rm{Mg})$, and this might
depend on uncertainties on the $\log gf$s and on the fact that
the analysis of this species is based on a very small number of lines.
The analysis of Na usually is based on a set of seven lines, but
in the case of NGC 6253 not all these features are measurable
and in addition the few lines used give discrepant results.
For this reason, the ratios of Na and Mg  to Fe in stars
of NGC 6253 have been computed
from a line-to-line comparison, instead of comparing the average
abundance of a star to the solar one.
The abundances of the two elements are presented separately in Table~\ref{line_to_line},
where we report [X/H] and [X/Fe] for each line
in common between the lists adopted for the Sun and for the cluster stars.

We mention that
for Mg only  two lines were used, and, as a consequence,
the results shown in Table~\ref{line_to_line},
which indicate a Mg enhancement,
should be taken with caution and would require a dedicated study
that goes beyond the primary goal of the present paper.
We used three spectral features
for the determination of Na abundance, but the two lines at 6154--6160 \AA~are likely
to be more reliable than the 5688 \AA~line, which is rather strong, and might
deserve more detailed damping computations.
Na abundances derived with MOOG are based on the assumption of LTE.
As well known, this assumption may introduce systematic errors
in the computation of the abundance;
whereas for most of the elements it has been ascertained that
non-LTE corrections are negligible,
in the case of Na they might be important.
The problem is that non-LTE effects affect at different levels
stars in diverse evolutionary phases (main sequence, TO, RGB, clump)
since they are strongly dependent on the temperature and surface gravity.
Also, discrepant results have been obtained from different authors
in the computation of non-LTE corrections.
For example, Gratton et al.~(\cite{gratton99})
find  moderate negative corrections of the order of
$\sim$0.05--0.1 dex for giant stars, while Mashonkina 
et al.~(\cite{masho}) estimate larger corrections, of the order of $\sim$0.15
dex. 
In Table~\ref{line_to_line} we show [Na/Fe] values derived with MOOG
and corrected adopting the tabulations by Mashonkina et al.
Considering LTE abundances, Na seems to be enhanced
with respect to the solar value, in agreement with other findings for open clusters
(Friel
et al.~\cite{friel03}; Yong, Carney, \& de Almeida \cite{yong05}; 
Bragaglia et al.~\cite{bragaglia06}), but a certain amount 
of scatter is present in
the abundances from the various lines and also among
the various stars. 
On the other hand, with the non-LTE corrections by Mashonkina et 
al.~the [Na/Fe]
values result to be lower by $\sim$0.10--0.15 dex depending on 
the line considered and
on the stellar parameters. In this case, the average Na abundance of the 
cluster would be nearly or slightly above solar;
therefore, we suggest that the Na abundance enhancement
claimed for open clusters based on giant stars might 
be in part related to non-LTE effects (see also Randich et al.~\cite{R06}).
On the other hand, the [Na/Fe] abundance ratios of field dwarfs
do not appear to be enhanced (e.g., Soubiran \& Girard \cite{sg05});
this issue
deserves further investigation, which is beyond the goals of
this paper.

\begin{table*}[!] \footnotesize
\caption{Elemental ratios ([X/H]) for stars in NGC~6253: Si, Ca, Ti, Ni and Ba.
Errors are the rms $\sigma_1$ -- due to
$EW$ uncertainties -- on [X/H]. The solar abundances found by us
are also shown ($\log n$(X)$_{\odot}$).}\label{alphaXH}
\begin{tabular}{ccccccccccccccccc}
\hline
\hline
Element     &069885          &023501          &023498 &024707 &105495 & Sun ($\log n$(X)$_{\odot}$)\\ 
\hline 
Si~{\sc i}  &+0.44$\pm$0.12 &+0.21$\pm$0.04 &+0.35$\pm$0.98   &+0.40$\pm$0.17   &+0.35$\pm$0.17 & 7.60$\pm$0.03\\
Ca~{\sc i}  &+0.50$\pm$0.08 & +0.32$\pm$0.12  &+0.32$\pm$0.10    &+0.35$\pm$0.12 &+0.25$\pm$0.10 & 6.35$\pm$0.04\\
Ti~{\sc i}  &+0.44$\pm$0.10 &+0.30$\pm$0.10   & +0.21$\pm$0.08&+0.59$\pm$0.11   & +0.34$\pm$0.08&  4.93$\pm$0.03\\
Cr~{\sc i}  &+0.45$\pm$0.09 &+0.29$\pm$0.07    &+0.36$\pm$0.11   & +0.49$\pm$0.17  & +0.33$\pm$0.15& 5.66$\pm$0.03\\
Ni~{\sc i}  &+0.44$\pm$0.09 & +0.34$\pm$0.09  & +0.38$\pm$0.08  & +0.50$\pm$0.09  &+0.67$\pm$0.11 &6.26$\pm$0.02 \\
Ba~{\sc ii} & +0.75$\pm$0.13& +0.40$\pm$0.15 & +0.63$\pm$0.14 &+0.57$\pm$0.16 & +0.73$\pm$0.04 & 2.13*\\
\hline
\end{tabular}

*The solar Ba abundance is from Anders \& Grevesse (1989). 
\end{table*}

\begin{table*}[!] \footnotesize
\caption{[X/Fe] abundances and averages.
Errors are the quadratic sum of $\sigma_1$ on [X/H] and on [X/Fe].}\label{alpha}
\begin{tabular}{ccccccccccccccccc}
\hline
\hline
Element     &069885          &023501          &023498 &024707 &105495 & Average1$\pm$rms& Average2$\pm$rms\\ 
\hline 
Si~{\sc i}  &$-$0.01$\pm$0.15 &$-$0.08$\pm$0.10 &+0.03$\pm$0.13   &+0.01$\pm$0.19   &+0.14$\pm$0.20 & +0.02$\pm$0.08 & $-$0.01$\pm$0.05\\ 
Ca~{\sc i}  &+0.05$\pm$0.12   & +0.03$\pm$0.15  &0.00$\pm$0.13    &$-$0.04$\pm$0.15 &$-$0.24$\pm$0.14 & $-$0.04$\pm$0.12 & +0.01$\pm$0.04\\ 
Ti~{\sc i}  &$-$0.01$\pm$0.13 &+0.01$\pm$0.13   & $-$0.11$\pm$0.14&+0.20$\pm$0.14   & $-$0.15$\pm$0.13&$-$0.01$\pm$0.14  & +0.02$\pm$0.13\\ 
Cr~{\sc i}  &+0.01$\pm$0.13   &0.00$\pm$0.11    &+0.04$\pm$0.24   & +0.10$\pm$0.19  & $-$0.16$\pm$0.18&$-$0.02$\pm$0.10  & +0.04$\pm$0.05\\ 
Ni~{\sc i}  &$-$0.01$\pm$0.13 & +0.05$\pm$0.15  & +0.06$\pm$0.12  & +0.11$\pm$0.13  &+0.18$\pm$0.15& 0.08$\pm$0.07 & +0.05$\pm$0.05\\ 
Ba~{\sc ii} & +0.30$\pm$0.16& +0.11$\pm$0.17 & +0.31$\pm$0.17 &+0.18$\pm$0.18 & +0.24$\pm$0.11& +0.23$\pm$0.08& +0.23$\pm$0.10\\
\hline
\end{tabular}

Averages: 1: computed including all the 5 stars; 2: computed excluding the clump star 105495.
\end{table*}

\begin{figure*}
\psfig{figure=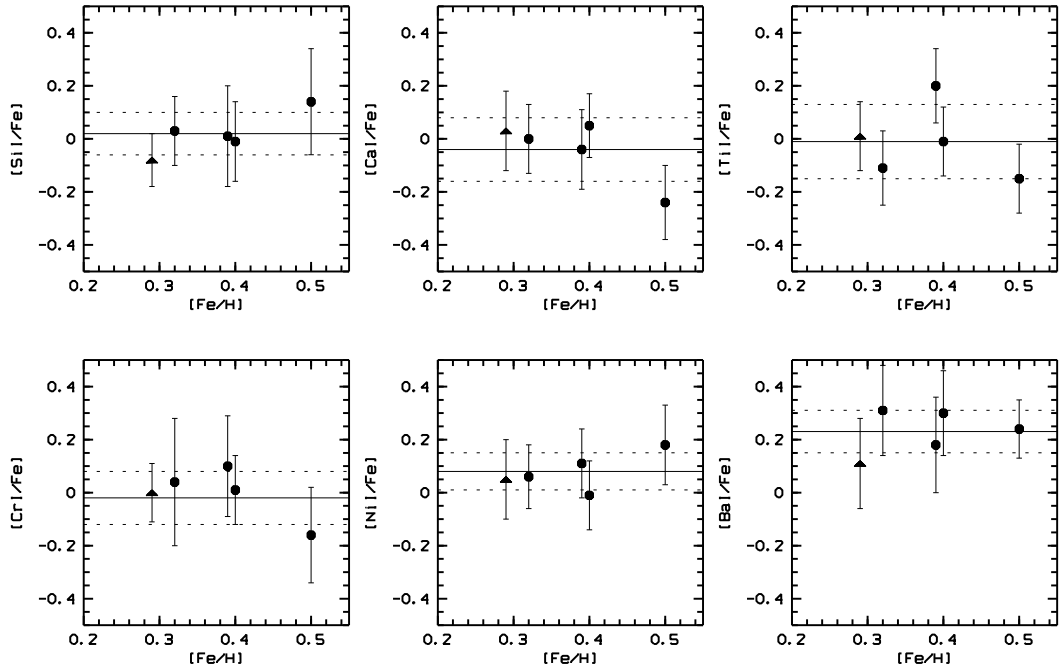, width=15cm, angle=-90}
\caption{[X/Fe] abundances of the various elements analyzed
as a function of [Fe/H]. The solid and dotted lines
are the averages $\pm$ rms (computed including
all the 5 stars plotted; Average1 in Table~\ref{alpha}); the triangle represents the possible member.}\label{elements_Fe}
\end{figure*}

We show in Fig.~\ref{elements_Fe} the element abundances of Table~\ref{alpha}
as a function of [Fe/H];
also in this case the solid 
and dotted lines represent the mean abundance
$\pm$ rms; we adopted the value Average1 shown Table~\ref{alpha},
i.e. computed including all the 5 stars.
All the elements, apart from  Ba, have average solar abundances;
Ba is enhanced,
as already found for other clusters, e.g., by our group (Bragaglia et al.
2007, in preparation)
or by Bragaglia et al.~(\cite{bragaglia06}); however, abundance for this
element has been found to vary significantly between clusters (e.g. Gratton, Sneden, \& Carretta \cite{G04}).

\begin{landscape}
\begin{table*}[!] \footnotesize
\caption{Abundances of Mg and Na (LTE and non-LTE) computed for each line adopted.}\label{line_to_line}
\begin{tabular}{cccccccccccccccccccc}
\hline
\hline
\scriptsize
&&\multicolumn{2}{c}{069885}&\multicolumn{2}{c}{023501}&\multicolumn{2}{c}{023498}&\multicolumn{2}{c}{024707}&\multicolumn{2}{c}{105495}\\
\hline
Wavelength& $\log n$(X)$_{\odot}$&[X/H]&[X/Fe]&[X/H]&[X/Fe]&[X/H]&[X/Fe]&[X/H]&[X/Fe]&[X/H]&[X/Fe]\\
\multicolumn{12}{c}{Mg~{\sc i}}\\
  6318.71      & 7.54&    +0.42   &$-$0.03&+0.71  & +0.42 &+0.75  & +0.43    & +0.72 & +0.33 & +0.86  & +0.37    \\
  6319.24      & 7.47&    +0.65   &+0.20  &--    & --    &+0.41  & +0.09    & +0.65 & +0.26 & +0.94  & +0.45     \\
Average      &     &             &+0.09$\pm$0.16    &         &+0.42 &      &+0.26$\pm$0.24   &  &+0.30$\pm$0.05   &    &+0.41$\pm$0.06     \\
\multicolumn{12}{c}{Na~{\sc i}}\\
   5688.22    & 6.20&    +0.69    & +0.24 &  +0.42    & +0.13 &  +0.77 & +0.45& +0.40 &+0.01  &+0.44    &$-$0.05	   \\
   6154.23    & 6.29&    +0.52    & +0.07 &  +0.51    & +0.22 &  +0.61 & +0.29& +0.75 &+0.36  &+0.88    &+0.39	    \\
   6160.75    & 6.30&    +0.60    & +0.15 &  +0.21    &$-$0.08&  +0.51 & +0.19& +0.69 &+0.30  &+0.77   &+0.28	    \\
Average      &     &             &+0.15$\pm$0.09    &         &+0.09$\pm$0.15 &      &+0.31$\pm$0.13   &  &+0.22$\pm$0.19   &    &+0.21$\pm$0.23      \\
Wavelength& &&[Na/Fe]$_{\rm non-LTE}$&&[Na/Fe]$_{\rm non-LTE}$&&[Na/Fe]$_{\rm non-LTE}$&&[Na/Fe]$_{\rm non-LTE}$&&[Na/Fe]$_{\rm non-LTE}$\\
   5688.22    & &        & +0.12 &     & +0.00 &  & +0.32& &$-$0.14  &    &$-$0.20	   \\
   6154.23    & &        & $-$0.05 &     & +0.10 &  & +0.17& &+0.27  &    &+0.24	    \\
   6160.75    & &        & +0.03 &     &$-$0.20&  & +0.02& &+0.21  &    &+0.13	    \\
Average      &     &             &+0.03$\pm$0.09    &         &$-$0.03$\pm$0.15 &      &+0.17$\pm$0.15   &  &+0.11$\pm$0.22   &    &+0.06$\pm$0.23      \\
\hline 

\hline
\end{tabular}
\end{table*}
\end{landscape}

\section{Discussion}\label{confronto}

\begin{figure*}
\psfig{figure=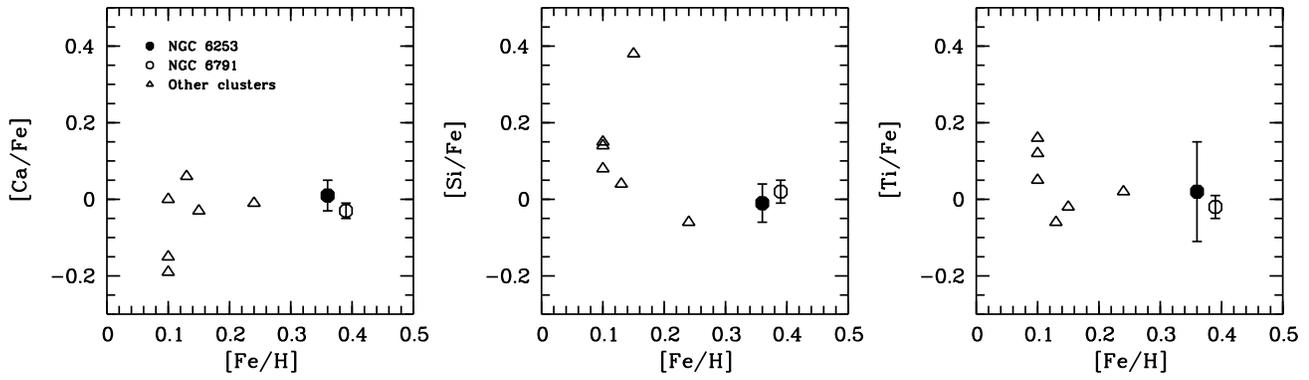, width=13cm, angle=-90}
\caption{$\alpha$-element abundances vs.~[Fe/H]: comparison
between NGC 6253 (filled circle), NGC 6791 (open circle, Carraro
et al. 2006), and other open clusters with oversolar [Fe/H]
(open triangles). 
For NGC 6253,
at variance with previous figures, 
we adopt here and in Figs.~\ref{fig_confr2}, \ref{fig_confr3}
the mean computed excluding the
clump star (called Average2 in Table~\ref{alpha}).}\label{fig_confr1}
\end{figure*}

\begin{figure*}
\psfig{figure=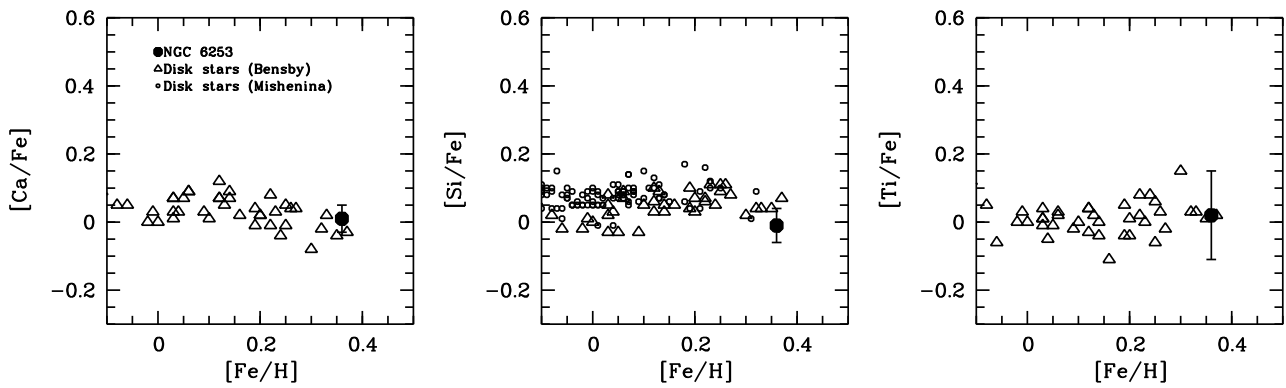, width=13cm, angle=-90}
\caption{$\alpha$-element abundances vs.~[Fe/H]: comparison
between NGC 6253 (filled circle) and disk stars (open
triangles, Bensby et al.~\cite{bensby} -- thick disk; open
circles Mishenina et al.~\cite{mishe04} -- thick and thin disk). In the latter study
Ca and Ti were not investigated.}\label{fig_confr2}
\end{figure*}

\begin{figure*}
\psfig{figure=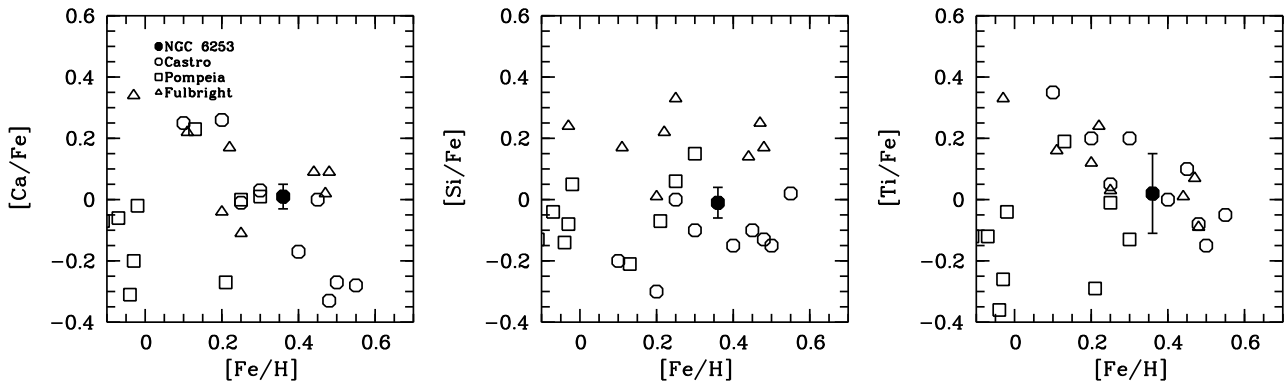, width=13cm, angle=-90}
\caption{$\alpha$-element abundances vs.~[Fe/H]: comparison
between NGC 6253
(filled circle), bulge stars
(open triangles, Fulbright,
McWilliam, \& Rich~\cite{fulb07}), and bulge-like field stars observed
by Castro et al.~(\cite{castro}, open circles) and
Pompeia et al.~(\cite{pompeia}, open squares).}\label{fig_confr3}
\end{figure*}

As mentioned in the introduction,
the metallicity of NGC~6253 
is unusual for disk stars,
that normally have abundances close to solar.
The only other cluster for which a very high Fe content has been reported
is the 9 Gyr old NGC~6791.
Carraro et al.~(\cite{carraro06}) and Origlia
et al.~(\cite{origlia}) found [Fe/H]=+0.39 and +0.35, respectively,
and solar $\alpha$-element abundances.
NGC~6791 is a peculiar cluster, since it is
very massive and has a very eccentric
orbit, therefore, it is different
from NGC 6253 in several aspects.
In particular,
NGC~6253 is much younger than NGC 6791
and much less massive;
the orbit of NGC 6253 has not been studied yet, but it could be
interesting to have information on it.
Given its position towards the
Galactic center, NGC 6253
could have  been born either towards the bulge, where
the metallicity is high, or in a region of the disk
where a particular metal enrichment occurred. 
In order to get insights on the origin of this cluster,
we show in Fig.~\ref{fig_confr1}
a comparison for $\alpha$-element
abundances (Si, Ca, Ti) vs. [Fe/H]
in NGC 6253 (filled circle), NGC 6791
(open square) and other open clusters with [Fe/H]$\gesssim$+0.10
(open triangles); the clusters are
the Hyades, 
\object{NGC 5822},
\object{IC 4651}, \object{IC 4725}, and
\object{NGC 6705}
(references can be found in Friel \cite{friel06} and
Randich \cite{R07}).
We plotted these three elements since their analysis is based 
on a rather large sample of lines with respect to the other species,
and they are all explosive nucleosynthesis $\alpha$-elements, i.e.
they originate from type II Supernova events.
Note that for NGC 6253 we consider here
(and in Figs.~\ref{fig_confr2}, \ref{fig_confr3}) the values labeled as
Average2  in Table~\ref{alpha}, excluding the
clump star; similarly, for Fe we adopted the value +0.36, again
excluding the clump star.
The figure shows that the average  $\alpha$-element abundances of
NGC 6791 and NGC 6253 are identical or in very good agreement.
$\alpha$-element abundances in open clusters with 
oversolar Fe content are in general
close to solar
(with the exception
of a Si enhancement in IC~4725, Luck et al.~\cite{luck}).

Figure \ref{fig_confr2} shows a comparison between NGC~6253 and disk dwarfs
observed by Mishenina et al.~(\cite{mishe04}; thin and thick disk,
open circles) 
and Bensby et al.~(\cite{bensby}; thick disk, open
triangles). Also in this case
the $\alpha$-elements Si, Ca and Ti were considered;
the
figure shows  the range in metallicity
[Fe/H]$\sim-$0.1$\div$+0.4. 
The average abundances of evolved stars
in NGC~6253 match very well the general
trend observed for thin and thick disk dwarfs.
Finally, we plot in Fig.~\ref{fig_confr3}
a comparison between NGC~6253 and 
the results for bulge giant stars recently analyzed by 
Fulbright et al. (\cite{fulb07}, open triangles)
and bulge-like field stars by
Castro et al. ~(\cite{castro}; open circles) and 
Pompeia et al.~(\cite{pompeia};
open squares). In the latter works
samples of nearby dwarfs
with kinematics and metallicity characteristics of  a probable inner
disk or bulge origin have been investigated.
The bulge and bulge-like
field star samples cover a [Fe/H] range $\sim-1.20\div+0.60$, but 
we show abundances only for [Fe/H] larger than $\sim$--0.1.
Bulge and bulge-like stars are characterized by $\alpha$-enhancement
at very low metallicities (not visible in the figure)
with
a decrease towards solar and oversolar metallicities, as shown
in the plot; 
note however the much
larger dispersion with respect to disk stars reported in Fig.~\ref{fig_confr2}:
indeed there are stars showing
enhanced $\alpha$-element abundances also at solar/oversolar metallicities.

The comparisons between NGC~6253 and field stars suggest that 
the abundance of the cluster is in good agreement with the trend
observed in the disk (Fig.~\ref{fig_confr2}).
From this evidence, 
we can speculate that NGC~6253 was born in the Galactic disk,
in a region where a larger than normal Fe enrichment occurred; on
 the other hand, 
we do not observe an enhancement of $\alpha$-elements with
respect to Fe. 

\begin{figure*}
\psfig{figure=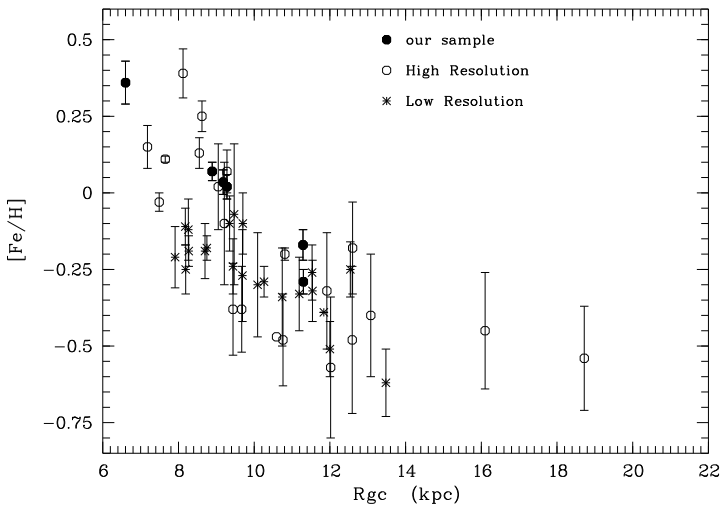, width=7cm, angle=-90}
\caption{Radial gradient ([Fe/H] vs.~Galactocentric distance) for open
clusters. The results for clusters
in our sample analyzed so far
(filled circles; this paper, Paper~{\sc i} and Bragaglia et al. 2007, in preparation) are compared to
other clusters analyzed with high-resolution spectroscopy (open circles,
see Paper~{\sc i} for references) 
and low-resolution spectroscopy (Friel et
al.~\cite{friel02}, asterisks).}\label{gradiente}
\end{figure*}

Under the assumption that the cluster formed in the disk,
we can use it for the determination of the radial metallicity
gradient.
In Fig.~\ref{gradiente} we show [Fe/H] as a function of the
Galactocentric radius for NGC~6253 and the other samples included in
our program (filled circles; Paper~{\sc i} and Bragaglia et al. 2007, in
preparation), 
compared to other samples analyzed 
with high resolution spectroscopy (open circles;
references for all the clusters investigated at high resolution can be found
in Paper~{\sc i}).
In the figure, we also show the low-resolution sample by
Friel et al.~(\cite{friel02}; asterisks; but note that we
excluded
the clusters in common with high resolution studies).
For consistency, we adopt for all the clusters $R_{\rm gc}$
from Friel (\cite{friel95}) and  Friel et al.~(\cite{friel02}).

NGC~6791 (the cluster with the highest [Fe/H])
lies above the average trend for open clusters,
confirming that it
might have had an origin and evolutionary history
very different from those of other clusters (Carraro et al.~\cite{carraro06}).
Also the very metal rich NGC 6253 (the other
cluster with [Fe/H]$>$+0.3) and
Praesepe ([Fe/H]=+0.25;
Pace et al. 2007, in preparation) might lie above the mean trend.
In other words, the inclusion of
NGC 6253 in the [Fe/H] vs. $R_{\rm gc}$
distribution makes the negative slope
of the gradient steeper. 
However, we note that all the clusters with $R_{\rm gc}$ lower than that
of the Sun -- $\sim$8.5 kpc -- have  higher
than solar metallicities, with the exception of Cr~261 which has [Fe/H]=$-$0.03
(Carretta et al.~\cite{carretta05}; De Silva et al. \cite{desilva}). This is true for
high resolution spectroscopy results, whereas Friel et al.~(2002),
from low resolution data, quote
metallicities from solar down to $-$0.25 dex for clusters
with  $R_{\rm gc}$ between 8 and 8.5 kpc.
Indeed,  the metallicities by Friel et al. from low
  resolution data are -- for all the clusters and at all $R_{\rm gc}$s
  -- lower than those obtained from high resolution
  analysis.
Considering only [Fe/H] values from high resolution and clusters 
with $R_{\rm gc}<$ 8.5 kpc,
we note that Cr~261 might represent an exception for its relatively low
Fe content, rather than NGC 6253 and other high metallicity clusters.
To our knowledge, no investigations of the orbit of
Cr 261 are present in the literature, but it would be very interesting
to have such an information, in order to understand if this cluster might have
formed at a larger Galactocentric
distance than its present position.

We tentatively conclude that the problems/questions raised by 
the very high metallicity of NGC 6791
should not really concern NGC 6253, 
which is rather young and
is located towards the Galactic center and very close to it
(at variance with NGC 6791 which is in the anticenter direction and
it is much older).
Finally, we would like to remark that, as mentioned in Sect.~1,
Twarog et al.~(\cite{ttl03}) suggested the possibility of an $\alpha$-element
enhancement in NGC 6253; we find instead solar-scaled abundances
for these elements,
implying that
also stellar evolutionary models at these metallicities need to be improved.

\section{Summary}\label{summary}

We report on chemical abundances in the
metal rich cluster NGC 6253, observed with  VLT/FLAMES.
The original sample includes seven stars (two at the
turn off, one at the clump, and four on the subgiant/red giant branch).

We find the following results:

\begin{enumerate}
\item Membership: among the seven stars in the original sample,
we considered
two objects along the RGB as non-members, since we are not able to
discern if their radial velocities are variable;
one of the two stars at TO is a radial velocity doubtful member, but
its metallicity is consistent with those of the members, therefore
we include it in the final sample. 
\item Metallicity: we confirm that the metallicity of NGC 6253 is much higher than
solar. In particular we find an average [Fe/H]=+0.39$\pm$0.08,
or [Fe/H]=+0.36$\pm$0.07 if we exclude the clump star.
\item Other elements: the abundance ratios of the $\alpha$-elements
Si, Ca, Ti and of the Fe-peak elements Cr and Ni
are solar in average, while Ba is enhanced, as usual for open clusters.
No trends with \teff~or [Fe/H] are present.
The clump star shows a Si-enhancement and Ca, Ti, Cr abundances
lower than solar;
nevertheless this result might be an artifact of the Fe scale offset.
The $\alpha$-element Mg appears to be enhanced in all the stars,
but the analysis is based only on two lines which give
scattering results. 
Also the abundance of the light element Na appears enhanced 
(at [Na/Fe] about +0.2) if
  the analysis is carried out in LTE. On the other hand, if the non-LTE
  corrections by Mashonkina et al.~(\cite{masho}) are taken into account, the
  average [Na/Fe] ratio becomes nearer to solar.
\item Origin of the cluster:
we compared the average abundances of 
the explosive nucleosynthesis $\alpha$-elements Si, Ca and Ti
in NGC 6253 (excluding the clump star),
and other stars of the Galactic population: open clusters, disk field
stars and bulge/bulge-like field stars.
\begin{itemize}
\item Comparisons with open clusters and field stars:
$\alpha$-element abundances in NGC 6253 are similar to
those of other open clusters and field disk stars
with oversolar Fe content; on the other hand, field stars located towards the
bulge are characterized by an $\alpha$-enhancement, with a large dispersion
for [Fe/H]$>$0.
Therefore,  it seems more likely that NGC 6253
was born in the disk, in a region where a particular Fe enrichment
occurred.
\item Fe gradient: in the context of the radial gradient
(the [Fe/H] distribution as a function of Galactocentric radius), 
NGC 6253 appears to lie slightly above the other clusters
with similar $R_{\rm gc}$, due to its
very high [Fe/H]. Therefore, either it had a different
formation history and evolution from other open clusters
(but this is unlikely for NGC 6253), or
the gradient is very steep
for clusters located at low $R_{\rm gc}$,
since the metallicities derived
from high resolution spectroscopy for almost all of them
are oversolar. 
\end{itemize}
\end{enumerate}

In summary, it is very difficult to speculate on the possible origin
of NGC 6253, although a very high Fe content has been ascertained.
Independently on its origin, 
this cluster is very interesting,
and it represents an ideal target where to search for extra-solar planets.

\begin{acknowledgements}
P.S. acknowledges support by the Italian MIUR, under PRIN 20040228979-001, and
by INAF-Osservatorio Astrofisico di Arcetri, where this work was completed.
We thank the anonymous referee for her/his valuable suggestions.
We are grateful to
 C.~Sneden and S.~Lucatello for having provided an updated version
of MOOG and for useful discussion on it; we  also thank
G. Carraro and L. Pasquini for helpful discussion and suggestions.
\end{acknowledgements}
{}

\end{document}